\documentclass[a4paper]{aa}
\usepackage{natbib}
\usepackage{epsfig}

\newcommand{\ylw}{\mbox{YLW\,15}}
\newcommand{\lhk}{\mbox{LkH$\alpha$\,92}}
\newcommand{\vtau}{\mbox{V773~Tau}}

\begin{document}

\title{Coronal structure geometries on pre-main sequence stars}

\author{F. Favata\inst{1} \and G. Micela\inst{2} \and F. Reale\inst{3}}

\institute{Astrophysics Division -- Space Science Department of ESA, ESTEC,
  Postbus 299, NL-2200 AG Noordwijk, The Netherlands
\and
Osservatorio Astronomico di Palermo, 
Piazza del Parlamento 1, I-90134 Palermo, Italy 
\and
Dip.\ Scienze FF.\ \& AA., Sez. Astronomia, Univ.\ Palermo,
Piazza del Parlamento 1, I-90134 Palermo, Italy
}

\offprints{F. Favata, \email{ffavata@astro.esa.int}}

\date{Received date / Accepted date}

\abstract{
  \textbf{ We have re-analyzed using a hydrodynamic model large
    flaring events on three different categories of pre-main sequence
    (PMS) stars: the young stellar object (YSO) \ylw, the classical T
    Tauri star (CTTS) \lhk, the weak-line T Tauri star (WTTS) \vtau,
    and the WTTS HD~283572 (the first three objects were observed by
    ASCA, the last by ROSAT; all have been previously reported in the
    literature).  The first three flares were previously analyzed on
    the basis of the quasi-static model mostly used up to now,
    consistently yielding large loops ($L \ga R_*$) and no evidence of
    sustained heating. Our hydrodynamic modeling approach, however,
    shows that the size of the flaring regions must be much smaller
    ($L \la R_*$) and moreover this method shows in all cases evidence
    of vigorous sustained heating during the flare decay, so that the
    decay of the observed light curve actually reflects the temporal
    profile of the heating rather than that of the free decay of the
    heated loop(s).  The events on the protostar \ylw\ have durations
    comparable to the stellar rotation period, so that their limited
    size and their lack of self-eclipses give evidence of a polar
    location on the star.  This is in contrast with the recently
    advanced hypothesis that these flares are due to long loops
    spanning the region between the star and the accretion disk. In
    general, the present analysis shows that flaring coronae on PMS
    stars have a structure similar to the coronae on older active
    stars.}  }

\maketitle 

\section{Introduction}
\label{sec:intro}

Strong X-ray emission is associated with all stages of the early
evolution of low-mass stars, from the protostellar Young Stellar
Objects -- YSO or Class~I -- to the Classical T~Tau -- CTTS or
Class~II -- to the Weak-Line T~Tau phase -- WTTS or Class~III. The
salient characteristics of stellar activity in the pre-main sequence
(PMS) phase have been recently reviewed by \citet{fm99}: in most cases
the observed X-ray emission has been interpreted as due to
magnetically confined (and likely magnetically heated) plasma,
although with much enhanced -- both in absolute and relative terms --
activity levels with respect to main-sequence stars of comparable
mass. Whether the corona is however just an enhanced version of a
``solar-like'' one, or whether significantly different mechanisms (and
coronal topologies) are at work has been a matter of debate, in
particular in relationship to the possible influence of the accretion
disk on the corona: in YSO's and CTTS's significant amounts of
circumstellar material (mostly in the form of a disk) is present, and
magnetic fields are thought to funnel and modulate the accretion onto
the (proto)star \citep[e.g.][]{har98}. Various possibilities for the
magnetic interaction between the star and the disk are for example
shown in Figs.~2 and~3 of \citet{fm99}; understanding the topology and
dimensional scales of the coronal plasma (and, indirectly, of the
magnetic field) can, in the case of PMS stars, give important
constraints to the accretion process. The stellar activity also has a
significant influence on the circumstellar environment \citep[see e.g.
the review of][]{gfm2000}, and for example a low-lying X-ray emitting
region will irradiate (and thus ionize) the circumstellar disk much
less than a very extended emitting region from which X-rays can reach
the disk out to much larger distances from the parent star.

The study of the decay of flares has in the last decade been
extensively used to model the characteristics of the flaring regions
in different stellar types, and in particular their size. This in turn
has been used to derive constraints on the structuring of the active
corona, with results which are however, as discussed below,
model-dependent.  The quasi-static method of flare modeling
\citep[originally described by][]{om89} \textbf{assumes that the
  flaring loops decay freely after an initial impulsive heating event,
  and that the decay can be described through a series of static
  states;} while in principle the formalism allows for the presence of
heating during the decay phase, in practice in no case has the
application of the formalism resulted in sustained heating being
detected. Thus, in the presence of the long decay times typically
observed for large flares in active stars, the quasi-static method
consistently yields long loops and extended coronae. This method has
been extensively applied in the last decade, resulting in a general
assumed framework of coronae of active stars populated with very long,
extended loop structures. In the case of PMS stars, these have often
been thought to extend to, and link with, the accretion disk,
resulting in a distinctly ``non-solar'' coronal geometry.

More recently, \citet{rbp+97} have developed a method based on
hydrodynamic modeling of decaying flaring loops, which is able to
detect the presence of heating during the decay and to adjust the
estimate of the loop length accordingly, and which has been
\textbf{tested (``calibrated'') with good results on solar flares of
  which the loop length can be checked from images}. When applied to
the only stellar flare for which an eclipse allows to make a geometric
size estimate \citep[the Algol SAX flare,][]{sf99, fs99}, the
quasi-static method has been shown to over-estimate the loop size by a
large factor, while the hydrodynamic modeling diagnoses the presence
of long-lasting sustained heating during the decay, and yields a more
realistic estimate of the region's size (although still overestimating
it).  Application of this method to a number of events on different
types of coronal sources has shown that sustained heating is an
essential feature of large stellar flares, and that therefore the
flaring structures are actually significantly smaller than so far
postulated.  This applies to flare stars (CM Leo, \citealp{rm98}; EV
Lac, \citealp{frm+2000}; AD Leo, \citealp{fmr2000}), to young single
solar-type stars (AB~Dor, \citealp{mpr+2000}), to active binaries (AR
Lac and CF Tuc, \citealp{fav2001}) and to Algol \citep{fs99,
  fmr+2000}.

The hydrodynamic modeling approach is based on detailed simulations of
decaying flaring loops. The simulations show that the key diagnostic
is the slope $\zeta$ of the flare's decay in the $\log T$ vs. $\log n$
plot (actually, in the case of stellar flares, where no direct density
diagnostics are usually available, the $\log T$ vs. $\log
\sqrt{E\!M}$\footnote{\textbf{where the emission measure is defined as
    $E\!M = \int{n_e n_{\rm H} dV}$}} plot): if sustained heating is
present, the slope will be shallower than in the case of an
impulsively heated, freely decaying loop \textbf{(cf. e.g.,
  Fig.~\ref{fig:ylw15})}.  A fit to the parameters of the numerical
model allows to derive the length of the flaring loop as a function of
the observed decay time scale $\tau_{\rm LC}$, of the $\zeta$
parameter and of the peak temperature in the flaring loop $T_{\rm
  max}$. \textbf{The latter quantity} is defined as the maximum
intrinsic temperature of a model loop whose spectrum synthesized in
the band of a given X-ray spectrometer yields, when fit with a
single-temperature spectrum, a maximum \emph{observed} temperature
$T_{\rm p}$.

Large flares on PMS stars have in the past always been studied with
the quasi-static method, resulting in typical sizes of $L \simeq
10^{11}$--$10^{12}$~cm, i.e.\ $L \ge R_*$, which have often lead to
the hypothesis of loops linking the photosphere with the accretion
disk (e.g. \citealp{tik+2000}). In the present paper we study, through
the hydrodynamic modeling method of \citet{rbp+97}, stellar X-ray
flares observed on PMS stars from Class~I to Class~III, with the aim
of determining whether they feature strong sustained heating (as
observed in all stellar types studied so far with this approach) and
thus whether the related coronal structures are significantly more
compact than postulated up to now. Given the relatively large average
distance of star-forming regions (and therefore of PMS stars) from
Earth, only few (and very intense) X-ray flares from PMS stars have
been observed with sufficient statistics, and therefore the choice of
the sample studied in this paper is driven by the (random)
availability of the data, and is neither complete or unbiased.

The present paper is so structured: Sect.~\ref{sec:sample} gives a
summary description of the sample's characteristics, in
Sect.~\ref{sec:anal} the analysis of each flare is discussed in
detail, while in Sect.~\ref{sec:disc} the results are discussed and
summarized.

\section{Sample studied}
\label{sec:sample}

We have identified, in the literature, four X-ray flares from PMS's
which have sufficient statistics to allow their study with the
hydrodynamic modeling approach: a flare observed with ASCA on the
protostar \ylw\ (\citealp{tik+2000}), a flare observed with the ROSAT
PSPC on the CTTS \lhk\ (\citealp{pzs93}), a flare observed with ASCA on
the WTTS \vtau\ (\citealp{tkm+98}) and a PSPC flare on the WTTS HD~283572
(\citealp{snh2000}). The sample (although small) spans the complete
pre-main sequence phase and covers all key evolutionary stages; the
somewhat older star AB~Dor, just arriving on the main sequence, has
been studied with the same approach by \citet{mpr+2000}, thus linking
the present sample with the subsequent evolutionary stage.

The first three events discussed here have been previously analyzed
with the quasi-static method, while the fourth has been reported by
\citet{snh2000} but not analyzed in detail. The quasi-static analysis
has resulted (as typical for large stellar flares) in large loop sizes
($L \simeq 3.5~R_*$ for \ylw, $L \simeq 0.6 R_*$ for \lhk, $L \simeq
1.3~R_*$ for \vtau).

\section{Analysis of the flaring events}
\label{sec:anal}

\subsection{\ylw}

\ylw\ is a low-mass Class~I protostar in the $\rho$~Oph star-forming
region ($d \simeq 165$~pc), with an estimated radius $R_* \simeq
4.2\,R_\odot = 2.9 \times 10^{11}$~cm (e.g. \citealp{mgt+2000}). Its
high X-ray activity was noticed through a large flare observed with
the ROSAT HRI (\citealp{gmf+97}); the lack of spectroscopic
capabilities of the HRI however prevents a detailed analysis of this
event. \ylw\ was later observed with ASCA, which detected three
consecutive significant flaring events, which are discussed and
analyzed by \citet{tik+2000}.  The first and most intense of the three
events has a decay time-scale $\tau_{\rm LC} = 31$~ks, a
\textbf{measured maximum temperature} $T_{\rm d} \simeq 64$~MK and a
peak X-ray luminosity of $\simeq 2 \times 10^{32}$~erg~s$^{-1}$.
\citet{tik+2000} have analyzed the event using the quasi-static
framework (assuming impulsive heating), deriving a
\textbf{semi-length} of the flaring loop $L \simeq 10^{12}~{\rm cm}
\simeq 3.5~R_*$, a density $n \simeq 5 \times 10^{10}$~cm$^{-3}$ and a
minimum value, for the magnetic field, of $B \simeq 150$~G. Similar
physical parameters are derived for the second and third event (which
have however lower statistics).

\begin{figure}[tbp]
  \begin{center} \leavevmode \epsfig{file=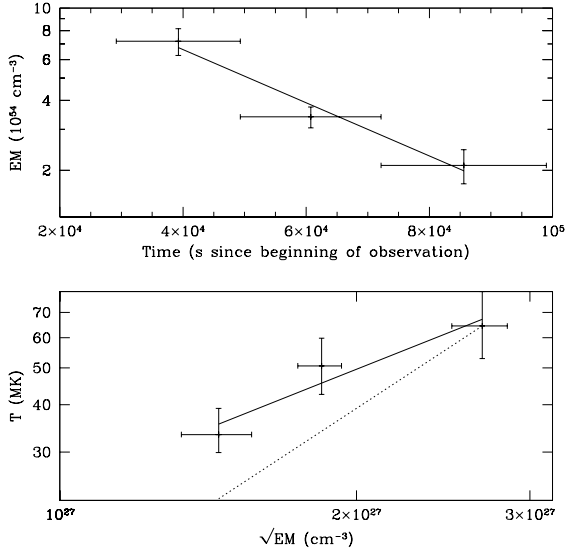, width=8.0cm,
  bbllx=10pt, bblly=150pt, bburx=600pt, bbury=700pt, clip=}

  \caption{The evolution of the \ylw\ flare. The top panel shows the
    evolution of the flare's emission measure (used as a proxy to the
    light curve, e.g. to the X-ray count rate), together with the
    best-fit exponential decay, while the bottom panel shows the
    evolution of the flare's decay in the $\log T$ vs.\ $\log n$
    plane. The solid line is the best fit to the decay ($\zeta =
    1.0$), while the dotted line is the path the decay would have
    followed if it had been quasi-static with no sustained heating
    during the decay phase ($\zeta_{QS} = 1.7$).}

  \label{fig:ylw15}
  \end{center}
\end{figure}

Using the spectral parameters ($T$ and $E\!M$) derived by
\citet{tik+2000} for three temporal intervals during the flare decay
we have modeled the first flare using the hydrodynamic modeling
approach of \citet{rbp+97}, recalibrated for use with the ASCA
detectors as described by \citet{frm+2000} and \citet{fmr2000}.
Although the flare decay has been divided in only three time intervals
(due to the limited number of photons detected), this is sufficient to
constrain the presence of sustained heating and thus the loop's
length. The flare evolution and its best-fit decay in the $\log T$
vs.\ $\log n$ plane are shown in Fig.~\ref{fig:ylw15}. The slope of
the decay, $\zeta = 1.0 \pm 0.3$ is significantly shallower than the
slope that would be observed if the decay were indeed quasi-static
($\zeta_{QS} = 1.7$). The corresponding ratio between the the observed
decay time and the intrinsic thermodynamic decay cooling time of the
loop without additional heating (\citealp{srj+91}) is $\tau_{\rm
  LC}/\tau_{\rm th} = F(\zeta) \simeq 2.7 [2.2 - 3.6]$\footnote{Here,
  as in the following, the notation $x [xl - xu]$ indicates the
  best-fit value of the quantity $x$ together with the lower and upper
  range of the $1 \sigma$ confidence interval, which is typically not
  symmetric.}, indicating the presence of significant sustained
heating during the decay. The actual peak temperature of the plasma in
the flaring loop is $T_{\rm max} \simeq 150$~MK, and the loop length
is $L \simeq 4.6 [3.5 - 5.6] \times 10^{11} {\rm cm} \simeq 1.6 [1.2 -
1.9]\,R_*$, a factor of $\simeq 2$ smaller than estimated by the
quasi-static method under the assumption of free decay.

To estimate the density $n$ of the plasma and the strength of the
confining magnetic field $B$, the volume $V$ of the flaring region and
therefore its geometry must be known. Assuming a loop with constant
cross-section, whose length is obtained through the flare analysis,
the only free parameter is the \textbf{ratio of the diameter of the
  loop's cross section to its total length}, or the aspect ratio,
$\beta$. In the solar context $\beta \simeq 0.1$ is often claimed as
``typical'', although in the case of active stars higher values (up to
$\beta \simeq 0.3$, \citealp{mpr+2000}; \citealp{fmr2000}) have been
inferred.

For a semicircular loop, $V = 2 \pi\ L^3 \beta^2$. For the \ylw\ flare
we thus derive $V = 6.1 \times 10^{33}$--$5.5 \times 10^{34}$ cm$^3$
(where the first value is for $\beta = 0.1$ and the second for $\beta
= 0.3$, as in the following). The density of the flaring plasma at the
peak of the event can then be derived as $n_e = \sqrt{E\!M/(0.8 V)}$,
where the factor 0.8 is the scaling factor between the number of ions
and the number of electrons for a solar H/He ratio. \textbf{This
  assumes a single loop uniformly filled with plasma. If the loop is
  significantly filamented (i.e. has a filling factor smaller than 1)
  the resulting density would be a lower limit to the actual value.}
In the present case (using the peak $E\!M = 7.2 \times 10^{54}$
cm$^{-3}$ reported by \citealp{tik+2000}), we derive $n_e = 3.8 \times
10^{10}$--$1.3 \times 10^{10}$ cm$^{-3}$. The plasma pressure can then
be estimated through $p_e = 2 n_e k T_p = 1600$--$500$ dyne cm$^{-2}$.
The equilibrium pressure estimated from the peak temperature and the
derived length of the flaring loop through the scaling laws of
\citet{rtv78} is $p_{\rm RTV} = 2700$ dyne cm$^{-2}$, showing that, in
this case, either the actual loop is quite thin, with $\beta < 0.1$,
or that the loop is significantly out of equilibrium, perhaps still
being filled with evaporating plasma when the decay starts. The
magnetic field necessary to confine the flaring plasma is $B_0 =
\sqrt{8 \pi p_e} = 200$--$120$ G. kG photospheric magnetic fields have
been observed through Zeeman splitting in flare stars by e.g.\ 
\citet{jv96}.

\subsection{The \vtau\ flare}

\vtau\ is a binary weak-line T~Tau star located in the Barnard 209
cloud (in the Tau-Aur star-forming region), at a distance $d \simeq
150$~pc. The two components have spectral types K2 and K5
(\citealp{wel95}), with an estimated radius for either component $R*
\simeq 4~R_\odot = 3 \times 10^{11}$~cm (\citealp{bck+95}). The
interbinary separation is $a \simeq 5 \times 10^{12}$~cm
(\citealp{wel95}).

A large X-ray flare detected during an ASCA observation of Barnard 209
has been studied by \citet{tkm+98}. The decay time-scale of the flare
is $\tau_{\rm LC} = 8.2$~ks with a \textbf{maximum observed
  temperature} in the ASCA-SIS spectrum of $T_{\rm d} \simeq 110$~MK
(i.e.  a very hot event). The peak X-ray luminosity is $L_{\rm X}
\simeq 10^{33}$~erg~s$^{-1}$. \citet{tkm+98} note that some of the
conditions for the applicability of a quasi-static analysis are
actually violated: in particular, the quantity $T^{3.25}/E\!M$, which
is supposed to be constant during the decay, increases toward the end
of the event, something which \citet{tkm+98} interpret as evidence for
reheating. They then use only the first part of the decay (in which
$T^{3.25}/E\!M$ is approximately constant) to derive the flare
parameters. Their quasi-static analysis results in a length of the
flaring loop $L \simeq 4 \times 10^{11}~{\rm cm} \simeq 1.3~R_*$ and a
density $n \simeq 3 \times 10^{11}$~cm$^{-3}$. The loop is thus
derived to be larger than the stars but much smaller than the
interbinary separation.

\begin{figure}[tbp]
  \begin{center}
    \leavevmode \epsfig{file=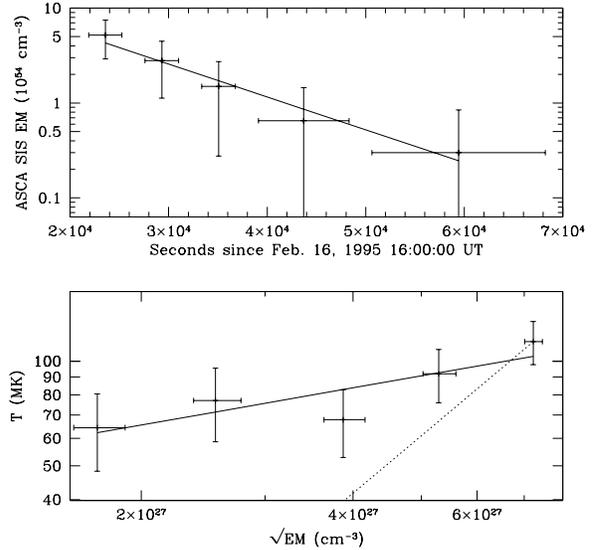, width=8.0cm, bbllx=10pt,
      bblly=150pt, bburx=600pt, bbury=700pt, clip=}
    \caption{The evolution of the \vtau\ flare. The top panel
      shows the flare's emission measure (used as a proxy to the light
      curve, e.g. to the X-ray count rate), together with the best-fit
      exponential decay, while the bottom panel shows the evolution of
      the flare's decay in the $\log T$ vs.\ $\log n$ plane. The solid
      line is the best fit to the decay, while the dotted line is the
      path the decay would have followed if it had been quasi-static
      with no sustained heating during the decay phase.}
    \label{fig:vtau}
  \end{center}
\end{figure}

Similarly to the other events discussed here, we have analyzed the
event using the hydrodynamic modeling, starting from the the spectral
analysis results of \citet{tkm+98}. The results are shown in
Fig.~\ref{fig:vtau}: the decay of the temperature during the flare is
quite shallow, so that the shape of the light curve is fully dominated
by the temporal evolution of the heating, with $\zeta = 0.36 \pm 0.12$
and $F(\zeta) = \tau_{\rm LC} / \tau_{\rm th} = 6.4 [5.3 - 7.8]$. The
actual peak temperature of the event is very high, at $T_{\rm max}
\simeq 200$~MK, and the loop length is $L \simeq 7.5 [6.1 - 9.0]
\times 10^{10}~{\rm cm} \simeq 0.25 \pm 0.05\,R_*$. The event,
notwithstanding its high temperature and X-ray luminosity, is
originating in a rather small loop, and therefore close to the stellar
surface.

Using the same approach as discussed for \ylw, the loop volume can be
estimated at $V=2.5 \times 10^{31}$--$2.3 \times 10^{32}$ cm$^3$, and
the density and pressure can be derived (using the peak $E\!M = 5.2
\times 10^{55}$ cm$^{-3}$) at $n_e = 1.6 \times 10^{12}$--$5.3 \times
10^{11}$ cm$^{-3}$ and $p_e = 2 N_e k T_p = 8.6\times
10^4$--$2.9\times 10^4$ dyne cm$^{-2}$. The equilibrium pressure is
$p_{\rm RTV} = 3.8\times 10^4$ dyne cm$^{-2}$, compatible with a loop
with $\beta \simeq 0.3$. The confining magnetic field is estimated at
$B_0 = 1500$--$850$ G.

\subsection{The \lhk\ flare}

\lhk\ is a classical T~Tau star in the IC~348 star-forming complex,
with a K spectral type, and an estimated radius $R_* \simeq 2~R_\odot
= 1.4 \times 10^{11}$~cm. A large X-ray flare has been detected in its
light curve during a ROSAT PSPC observation of IC~348, which has been
analyzed by \citet{pzs93} using the quasi-static approach. The
analysis of \citet{pzs93} resulted in a decay time-scale of the flare
$\tau_{\rm LC} = 7.8$~ks, and a observed peak temperature determined
from the PSPC spectra $T_{\rm p} \simeq 43$~MK. The peak X-ray
luminosity is $\simeq 5 \times 10^{32}$ erg~s$^{-1}$.  The
quasi-static analysis results in a length of the flaring loop $L
\simeq 8 \times 10^{10}~{\rm cm} \simeq 0.6~R_*$ (i.e.\ a
moderate-size loop) and a density $n \simeq 1.5 \times
10^{11}$~cm$^{-3}$.

\begin{figure}[tbp]
  \begin{center}
    \leavevmode \epsfig{file=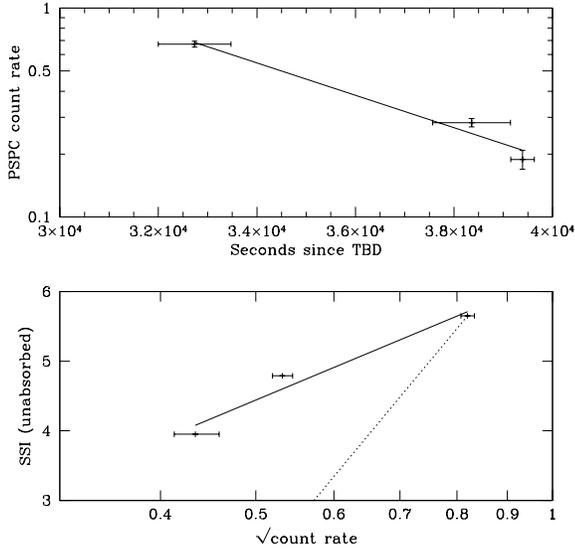, width=8.0cm, bbllx=10pt,
      bblly=150pt, bburx=600pt, bbury=700pt, clip=}
    \caption{The evolution of the \lhk\ flare. The top panel
      shows the flare's background-subtracted light curve together
      with the best-fit exponential decay, while the bottom panel
      shows the evolution of the flare's decay in the SSI (Spectral
      Shape Index) vs.\ $\log n$ plane (using the square root of the
      count rate as a proxy to the density $n$). The solid line is the
      best fit to the decay, while the dotted line is the path the
      decay would have followed if it had been quasi-static with no
      sustained heating during the decay phase.}
    \label{fig:lhk}
  \end{center}
\end{figure}

We have analyzed the event using the approach discussed in
\citet{rm98}, based on a principal component analysis, allowing to
make optimal use of the low-resolution, limited bandpass spectra
produced by the ROSAT PSPC. The key feature of this approach is the
use of a non-parametric temperature indicator (the Spectral Shape
Index, or SSI), rather than a parametric fit to the spectrum.

For the analysis, the flare decay has been binned into three
intervals, as shown in Fig.~\ref{fig:lhk}, extracting
background-subtracted spectra for each of them. The quiescent emission
(determined in the pre-flare phase) was also subtracted from the flare
spectra. The observed decay time scale for the event is $\tau_{\rm LC}
= 5.6 [4.8 - 6.8]$ ks. The measured SSI values have been converted to
the intrinsic (unabsorbed) SSI values (as discussed in \citealp{rm98})
assuming an absorbing column density $N(H) = 10^{22}$ cm$^{-2}$ (as
derived by \citealp{pzs93}). The slope in the count-rate vs. SSI plot is
$\zeta^\prime = 5.9 \pm 1.2$.  The peak SSI corresponds to a maximum
temperature for the event $T_{\rm max} = 110$ MK, and the
corresponding loop length is $L = (3.6 \pm 1.1) \times 10^{10}~{\rm
  cm} = (0.26 \pm 0.08) R_*$, where the relatively uncertainty is
estimated according to the value reported in Table~2 of \citet{rm98}.
This flare is therefore confined to a low-lying loop, close to the
stellar surface, and with a size falling within the range observed for
solar flares. The heating time scale $\tau_{\rm H}$ is estimated (from
Fig.~3 of \citealp{rm98}) to be $\tau_{\rm H} \simeq 2 \times \tau_{\rm
  LC}$. The observed flare decay is again dominated by the sustained
heating. Note that the peak temperature implied by the peak SSI value
is, at $110 $MK, significantly higher than the maximum temperature
measured through spectral fitting ($43$ MK, see above). This is due to
the relatively soft bandpass of the PSPC detector.

Along the same lines discussed above for the \ylw\ event the loop
volume can be estimated at $V=2.9 \times 10^{30}$--$2.6 \times
10^{31}$ cm$^3$ (for $\beta = 0.1$ and $\beta = 0.3$ respectively),
and the density and pressure can be derived (using the peak $E\!M =
4.8 \times 10^{54}$ cm$^{-3}$) at $n_e = 1.4 \times 10^{12}$--$4.8
\times 10^{11}$ cm$^{-3}$ and $p_e = 2 N_e k T_p = 4.2\times
10^4$--$1.4\times 10^4$ dyne cm$^{-2}$. The equilibrium pressure is
$p_{\rm RTV} = 1.2\times 10^4$ dyne cm$^{-2}$, compatible with a
$\beta = 0.3$ loop near equilibrium (i.e. filled with plasma).  The
confining magnetic field is estimated at $B_0 = 1000$--$600$ G.

\subsection{The HD 283572 flare}

HD 283572 is a relatively nearby WTTS in the Tau-Aur star-forming
region, at a (Hipparcos-determined) distance of 128\,pc. Its optical
characteristics are described in detail by \citet{wbl+87}, who derived
an MK spectral type of G5\,IV, and a radius of $3.3\,R_\odot$, at an
assumed distance of 160\,pc, which becomes $2.7\,R_\odot = 1.9 \times
10^{11} {\rm cm}$ at the Hipparcos distance.  Its X-ray
characteristics have been studied in detail by \citet{fms98}. Thanks
to its high quiescent X-ray luminosity ($L_{\rm X} \simeq 2 \times
10^{31}$ erg\,s$^{-1}$) it is a well studied object and it has been
observed with most X-ray telescopes to date.

A large flare seen with the ROSAT PSPC is reported (but not discussed
in detail) by \citet{snh2000}. We have analyzed the event using the
same approach based on a principal component analysis as used above
for \lhk; note that none of the other flares on PMS flares reported by
\citet{snh2000} has a sufficient number of events to allow a similar
analysis.

\begin{figure}[tbp]
  \begin{center}
    \leavevmode \epsfig{file=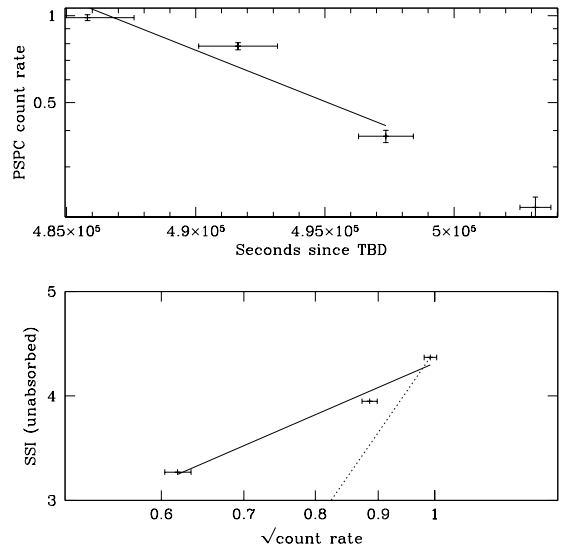, width=8.0cm, bbllx=10pt,
      bblly=150pt, bburx=600pt, bbury=700pt, clip=}
    \caption{The evolution of the HD~283572 flare. The top panel
      shows the flare's background-subtracted light curve together
      with the best-fit exponential decay, while the bottom panel
      shows the evolution of the flare's decay in the SSI vs.\ $\log
      n$ plane (using the square root of the count rate as a proxy to
      the density $n$). The solid line is the best fit to the decay.}
    \label{fig:hd}
  \end{center}
\end{figure}

The flare decay has been binned into four intervals, as shown in
Fig.~\ref{fig:hd}, extracting background-subtracted spectra for each
of them and subtracting the quiescent emission from the flare spectra.
Only the first three intervals have been used for the analysis, given
the very low statistics of the fourth interval. The observed decay
time scale for the event is $\tau_{\rm LC} = 12. [9.4 - 18.]$ ks; the
measured SSI values have been converted to the intrinsic (unabsorbed)
SSI values assuming an absorbing column density $N(H) = 10^{21}$
cm$^{-2}$ (as derived e.g.\ by \citealp{fms98}). The slope in the
count-rate vs. SSI plot of $\zeta^\prime = 5.1 \pm 0.8$. The peak SSI
corresponds to a maximum temperature for the event $T_{\rm max} = 48$
MK, with a corresponding loop length of $L = (5.0 \pm 1.5) \times
10^{10}~{\rm cm} = (0.26 \pm 0.08) R_*$. The relatively uncertainty is
also in this case is estimated according to the value reported in
Table~2 of \citet{rm98}. This flare is therefore also confined in a
low-lying loop, close to the stellar surface. The heating time scale
$\tau_{\rm H}$ is estimated (from Fig.~3 of \citealp{rm98}) to be
$\tau_{\rm H} \simeq 2 \times \tau_{\rm LC} \simeq 23.4$~ks.

The loop volume can be estimated at $V=7.8 \times 10^{30}$--$70 \times
10^{31}$ cm$^3$ (for $\beta = 0.1$ and $\beta = 0.3$ respectively),
and the density and pressure can be derived (using the peak $E\!M =
4.3 \times 10^{54}$ cm$^{-3}$) at $n_e = 8.3 \times 10^{11}$--$2.8
\times 10^{11}$ cm$^{-3}$ and $p_e = 2 N_e k T_p = 1.1\times
10^4$--$3.7\times 10^3$ dyne cm$^{-2}$. The equilibrium pressure is
significantly lower, $p_{\rm RTV} = 0.8\times 10^3$ dyne cm$^{-2}$,
thus implying either a loop with a very high aspect ratio $\beta >
0.3$ (which would however appear unlikely), or to the Spectral Shape
Index underestimating the peak temperature of the event: the
uncertainty on the SSI derived for a spectrum of about 1000 counts (as
it is the case for the hotter spectrum in the HD~283572 event) is,
again from Table~2 of \citet{rm98}, $\Delta_{\rm SSI} \simeq 0.5$, so
that the $1 \sigma$ interval for the event's peak temperature is $30$
-- $ 75$~MK, resulting in a $1 \sigma$ interval for the equilibrium
pressure $p_{\rm RTV} = [0.2 3.2]\times 10^3$ dyne cm$^{-2}$, so that
the upper value of the $1 \sigma$ interval is compatible with the
$\beta = 0.3$ configuration. Also, as shown by the other events
analyzed here, large flares on PMS stars tend to be quite hot, so that
indeed the nominal peak temperature of $48$~MK is lower than expected.
The confining magnetic field is estimated at $B_0 = 530$--$300$ G.

\section{Discussion}
\label{sec:disc}

\begin{table*}[tbp]
  \begin{center}
    \caption{A comparison of the physical parameters derived for
      the flaring events on PMS stars analyzed here. $T_{\rm d}$ is
      the maximum temperature measured during the flare (indicated
      only for the events analyzed through a direct spectral fitting,
      and absent for the events analyzed with the Spectral Shape Index
      -- SSI -- analysis), and $E\!M$ the maximum emission measure,
      while $L_{\rm X}$ is the peak X-ray luminosity. $\tau_{\rm LC}$
      is the decay time for the flare, while $L^{\rm QS}$ is the
      length of the flaring loop derived by quasi-static analysis,
      both in cm and in units of the stellar radius. $F(\zeta)$ is the
      ratio $\tau_{\rm LC}/\tau_{\rm th}$ between the observed decay
      time and the natural thermodynamic decay time of the loop
      (giving a measure of the importance of sustained heating in
      determining the light curve of the decay). $T_{\rm p}$ is the
      actual peak temperature in the flare derived from the observed
      temperature $T_{\rm d}$ and the instrument's response or from
      the value of the SSI, while $L^{\rm HM}$ is the length of the
      flaring loop derived through the hydrodynamic modeling approach
      (again in cm and in units of the stellar radius).  Numerical
      subscripts indicate the power of 10 by which the relevant
      quantity has been scaled.} \leavevmode
    \begin{tabular}{lrrrrrrrrr}
      Star & Instr. & $T_{\rm d}$ &$E\!M_{54}$ &
      $\tau_{\rm LC}$ & $L^{\rm QS}_{10}$ &
      $n_{10}^{\rm QS}$ & $F(\zeta)$ & $T_{\rm p}$ 
      & $L^{\rm HM}_{10}$ \\
      & & MK & \,cm$^{-3}$  & ks & cm ($R_*$) &
      cm$^{-3}$ & & MK &  cm ($R_*$) \\\hline
\ylw\ & ASCA-GIS &  64 & 15 & 37.9&100 (3.5) &  5 & 2.7 & 150 & 46.0
(1.5) \\ 
\lhk & ROSAT-PSPC&  43 & 50 & 5.6 &  8 (0.6) & 15 & 4.3 & 110 & 3.6 (0.3)
\\
\vtau & ASCA-GIS & 110 & 100 & 8.2 & 40 (1.3) & 30 & 6.4 & 200 & 7.5
(0.3) \\
HD~283572 & ROSAT-PSPC & -- & 4 & 12.0 & -- & -- & 4.6 & 48 &
5.0 (0.3) \\\hline
    \end{tabular}
    \label{tab:comp}
  \end{center}
\end{table*}

\begin{table}[tbp]
  \begin{center}
    \caption{The plasma physical parameters derived for the flaring
      events studied here. For each event, the density $n_e$, the
      pressure $p$ and the magnetic field $B$ needed to confine the
      plasma are tabulated under two different assumptions for the
      loop's aspect (length to diameter) ratio, i.e. $\beta=0.1$ and
      $\beta = 0.3$. Also, the ``equilibrium'' pressure $p_{\rm RTV}$
      that the plasma would have in a loop satisfying the scaling laws
      of Rosner et al.\ (1978) is reported. } \leavevmode
    \begin{tabular}{lrrrrr}
      Star & $\beta$ & $n_e$ & $p$ & $p_{\rm RTV}$ &  $B$  \\
      & & $10^9$ cm$^{-3}$&
      \multicolumn{2}{c}{$10^2$ dyne cm$^{-2}$} & kG \\\hline 
\ylw\ & 0.1 & 38 & 16 & 27 & 0.2  \\ 
 & 0.3 & 13 & 5 & 27  & 0.1  \\ \hline
\lhk & 0.1 & 1400 & 420 & 120 & 1.0 \\
 & 0.3 & 480  & 140 & 120 & 0.6 \\\hline
\vtau & 0.1 & 1600 & 860 & 380 & 1.5 \\
 & 0.3 & 530  & 290 & 380 & 0.9 \\\hline
HD~283572 & 0.1 & 830 & 110 & 8 & 0.5 \\
 & 0.3 & 280 & 37 & 8  & 0.3 \\\hline
    \end{tabular}
    \label{tab:par}
  \end{center}
\end{table}

All flares analyzed in the present paper show evidence for strong
sustained heating during the flare decay, so that the decay light
curve is dominated by the temporal heating profile and not by the free
decay of the heated loop. As a consequence, the decay light curve has
limited diagnostic value for the derivation of the physical parameters
of the flaring loop, and the quasi-static analysis which has been
previously applied to the events discussed here invariably results in
loop sizes which are significantly over-estimated. In the presence of
such vigorous sustained heating during the decay phase, even the sizes
derived by hydrodynamic modeling (which rely on a parameterization of
the heating temporal dependence as a decaying exponential) are likely
to result in upper limits to the actual loop sizes (as demonstrated by
the Algol SAX flare, see \citealp{fav2001} for a discussion). Therefore,
the large loop sizes obtained by quasi-static analysis ($L \ga 1 ~
R_*$) are in reality likely to be significantly smaller, typically by
factors $\ga 3$.

This result is in line with the results obtained by applying the same
hydrodynamic modeling approach to other classes of active stars, i.e.
on Algol (\citealp{fs99}; \citealp{fmr+2000}), on flare stars
(\citealp{rm98}; \citealp{fmr2000}; \citealp{frm+2000}), on the
zero-age main-sequence star AB Dor (\citealp{mpr+2000}) and on active
binaries (\citealp{fav2001}).  Sustained heating appears to be
consistently present during large X-ray flares, so that the flaring
regions are much smaller in size than it would be inferred by assuming
free decay of their light curve.  Therefore the (flaring) corona in
very active stars appears to be rather low-lying, close to the
photosphere, with no evidence for the very large, extended loops which
have been deduced in the past from the quasi-static analysis of large
flares. Pre-main sequence stars are no exception to this, as also all
flaring events analyzed here appear to be due to low-lying structures.

As discussed by \citet{fmr+2000}, the location of large flaring events
appears in general to be compatible with their being located at high
stellar latitude, close to the stellar poles, with no evidence for
low-latitude, near-equatorial structures (as for example typically
present on the Sun). No large flare has been observed to self-eclipse
even when its duration is comparable to the stellar rotational period;
given the small size of the flaring regions, the polar region is thus
the only plausible location for the flaring plasma. For the events
discussed here, the ones on \lhk, \vtau\ and HD~283572 have durations
significantly shorter than the plausible rotational period of the
stars and therefore this line of reasoning cannot be used to argue for
a polar location of the flaring region. For the event on \ylw,
\citet{mgt+2000} argue that the object is a fast rotator, with a
period $P_{\rm rot} = 16$--$20$~hr.  All three events reported by
\citet{tik+2000} have a duration of $\simeq 20$~hr each, and in no
case there is evidence for self-eclipse. Given the relatively small
size derived here for the flaring region, a polar location for the
flaring region on \ylw\ appears strongly favored.

This scenario (a low-lying flaring loop, likely located on the polar
region of the star) is not compatible with the explanation for the
large flares on \ylw\ proposed by \citet{mgt+2000}, who, on the basis
of the long loop size derived from the quasi-static analysis argue for
a flaring loop with one foot anchored on the stellar surface and the
other onto the (edge of the) accretion disk. The smaller loop size for
the \ylw\ flare implied by the hydrodynamic modeling discussed in the
present paper makes such loops implausible, and therefore it also
casts a doubt on the framework used by \citet{mgt+2000} to explain the
repeated flaring events seen on \ylw\ (i.e. the stressing of magnetic
field lines spanning both the star and the disc).

A summary of the characteristics of the event for each of the flares
studied here is reported in Table~\ref{tab:comp}, and a summary of the
resulting plasma parameters in Table~\ref{tab:par}. For the events on
the CCTS \lhk\ and on the WTTS \vtau\ and HD~283572 the resulting loop
sizes are consistently around $L \simeq 0.3 \, R_*$. The event on the
protostar \ylw\ is somewhat larger, with a nominal length $L \simeq
1.5 \, R_*$. However, the pressure derived even assuming $\beta = 0.1$
is significantly lower than the equilibrium pressure derived from the
\citet{rtv78} scaling laws. This points toward the loop being far from
equilibrium and only partially filled with hot plasma, so that the
procedure used here is likely to over-estimate the actual loop size.
Both these elements make us consider the length estimate derived here
as an upper limit to the actual size of the flaring region. Therefore
the characteristics of the X-ray flare observed on \ylw\ appear to be
similar to the ones observed on older stars, so that there may not be
any need to invocate a separate mechanism to explain the observed
behavior. Whether flares on protostars are in general more ``out of
equilibrium'' than flares on older stars is of course impossible to
say on the basis of a single event, and clearly study of a significant
number of flaring events will be needed to assess whether any such
general difference exists.

\begin{acknowledgements}

  FR, GM acknowledge the partial support of ASI and MURST.

\end{acknowledgements}


\end{document}